%% file: main.tex
\documentclass[journal=ancac3,manuscript=article]{achemso}

\usepackage{chemformula} 
\usepackage[T1]{fontenc} 
\usepackage{booktabs}
\usepackage{siunitx}
\DeclareSIUnit{\molar}{M}
\usepackage{hyperref}
\usepackage{gensymb}
\usepackage{tikz}
\usetikzlibrary{positioning, fit, patterns}

\author{Julian Ho{\ss}bach}
\altaffiliation{Equal Contribution}
\affiliation[University of Stuttgart]
{Institute for Computational Physics, University of Stuttgart, 70569 Stuttgart, Germany}
\author{Samuel Tovey}
\altaffiliation{Equal Contribution}
\affiliation[University of Stuttgart]
{Institute for Computational Physics, University of Stuttgart, 70569 Stuttgart, Germany}
\author{Tobias Ensslen}
\affiliation[University of Freiburg]
{Laboratory for Membrane Physiology and Technology, Department of Physiology, Faculty of Medicine, University of Freiburg, 79104 Freiburg, Germany.}
\author{Jan C. Behrends}
\affiliation[University of Freiburg]
{Laboratory for Membrane Physiology and Technology, Department of Physiology, Faculty of Medicine, University of Freiburg, 79104 Freiburg, Germany.}
\author{Christian Holm}
\affiliation[University of Stuttgart]
{Institute for Computational Physics, University of Stuttgart, 70569 Stuttgart, Germany}
\email{holm@icp.uni-stuttgart.de}
\title[Peptide Classification]
  {Peptide Classification from Statistical Analysis of Nanopore Translocation Experiments}

\keywords{Peptide characterization, Nanopore translocation, Machine learning}

\begin{document}


\begin{abstract}
  \input{content/abstract}
\end{abstract}

\input{content/introduction}
\input{content/results-and-discussion}
\input{content/conclusion}
\input{content/methods}

\begin{acknowledgement}
This work was supported by the BMBF FKZ 03ZU1208AM: nanodiag BW.
The authors of C.H.'s group acknowledge financial support from the German Funding Agency (Deutsche Forschungsgemeinschaft DFG) under Germany's Excellence Strategy EXC 2075-390740016.
Partially, this work was also supported by the Deutsche Forschungsgemeinschaft (DFG, German Research Foundation), Project-No 497249646, SPP 2363- "Utilization and Development of Machine Learning for Molecular Applications – Molecular Machine Learning.".
Initial work on this project in the lab of JCB was also funded by the BMBF through project ProNanoPep (PTJ 031B1120A).
\end{acknowledgement}

\begin{suppinfo}
The supporting information includes:

\begin{itemize}
    \item t-SNE plots for different parameters.
    \item Plots of the fitting process.
    \item Extended description of the catch22 features.
    \item Details on the nanopore fabrication and experimental data collection.
\end{itemize}

\end{suppinfo}

\bibliography{bibliography}

\end{document}


\section{TSNE Embeddings}

\begin{figure}[H]
    \includegraphics[]{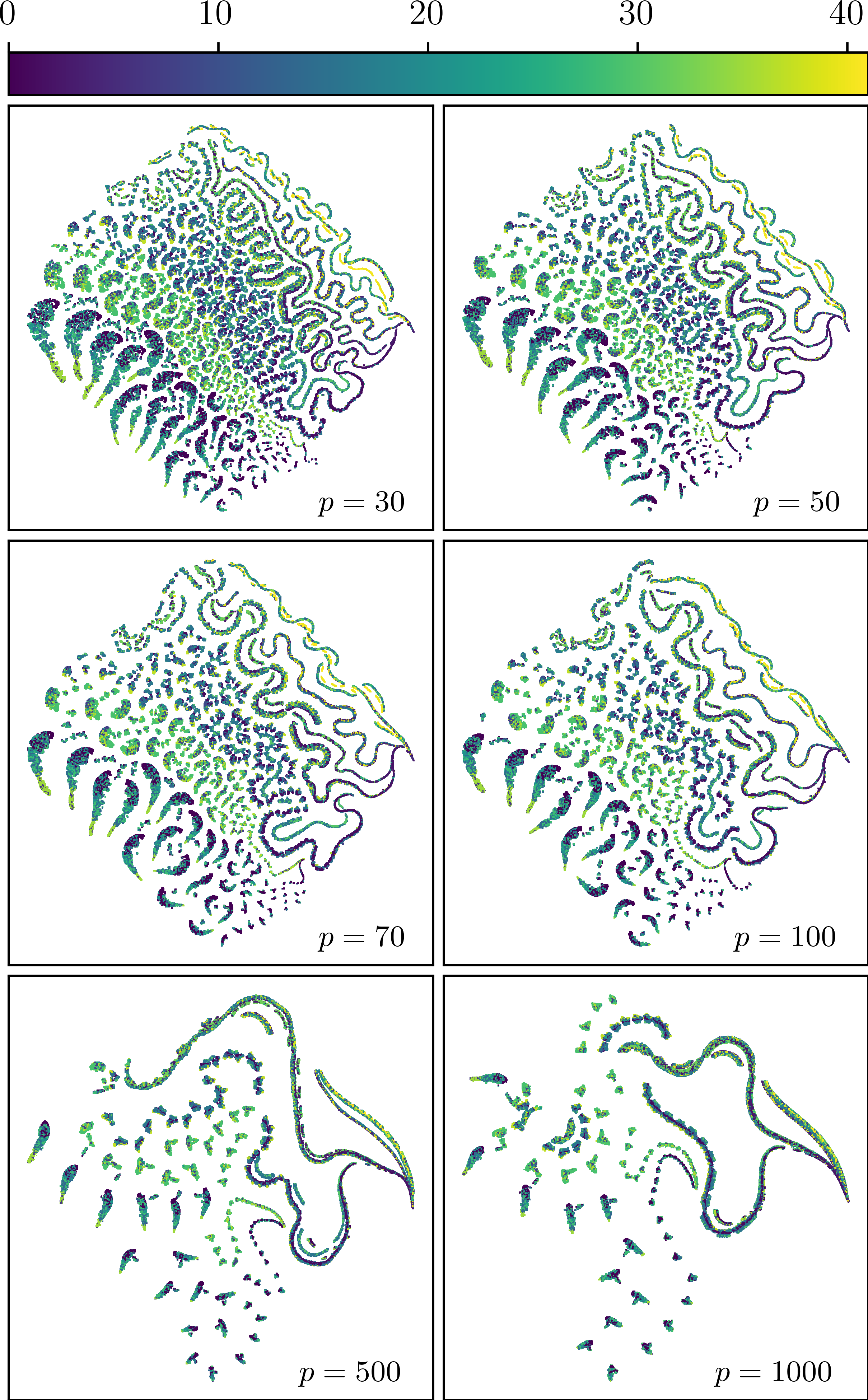}
    \caption{t-SNE images of the statistical moment decomposition feature set. The color denotes the actual labels corresponding to the different peptides. $p$ denotes the perplexity used in the algorithm.}
\end{figure}

\begin{figure}[H]
    \includegraphics[]{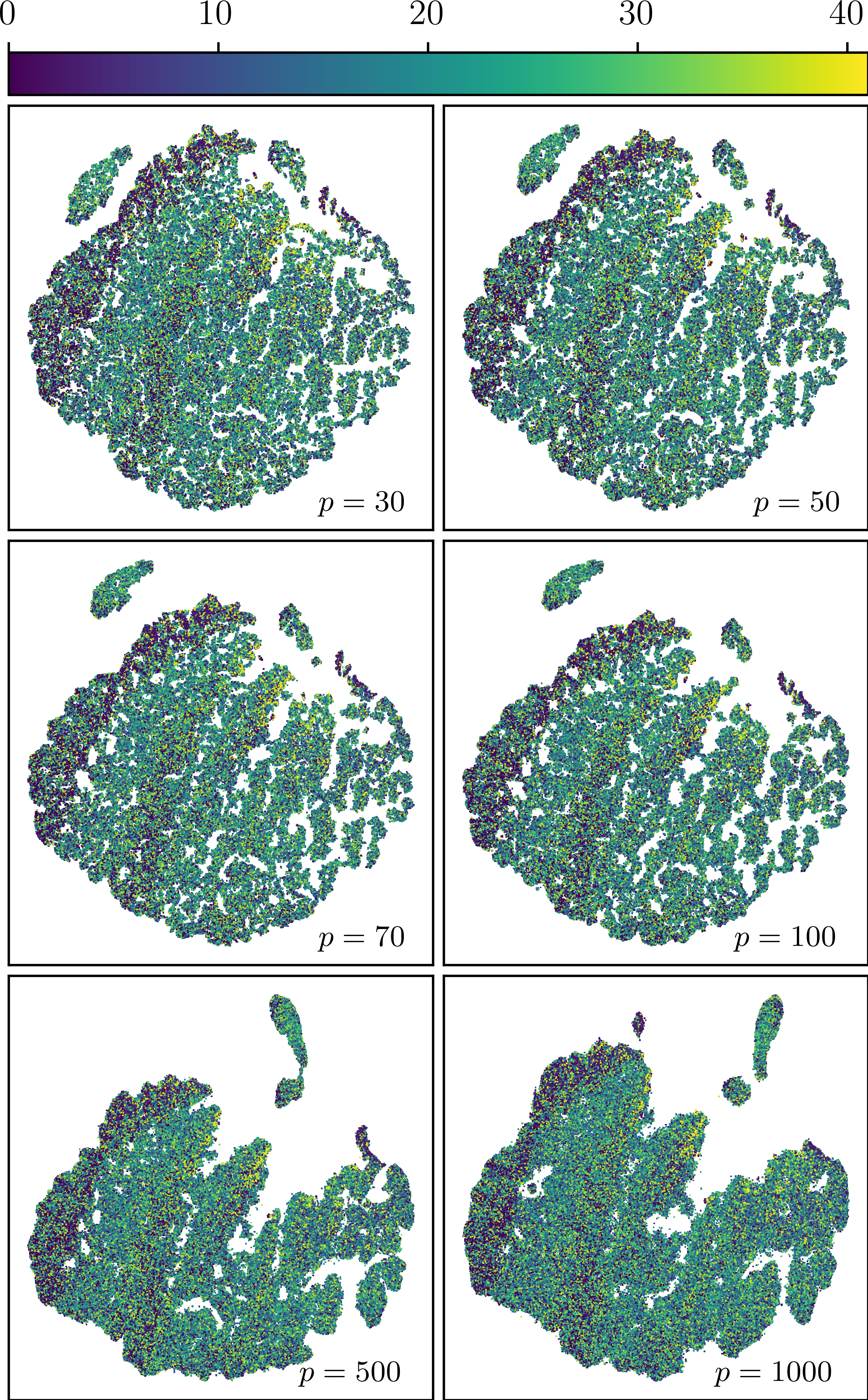}
    \caption{t-SNE images of the catch22 feature set. The color denotes the actual labels corresponding to the different peptides. $p$ denotes the perplexity used in the algorithm.}
\end{figure}

\section{Labels and corresponding peptides}

\begin{table}[H]
    \centering
    \caption{Label used in the classifier and the corresponding peptide.}
    \begin{tabular}{|c|l|l||c|l|l|}
        \hline
        Label & Peptide & Sequence & Label & Peptide & Sequence \\
        \hline
        0  & L1AS4  & H-RRRR-OH       & 21 & L4AS4  & H-KRRR-OH        \\
        1  & L1AS5  & H-YRRRR-OH      & 22 & L4AS5  & H-SKRRR-OH       \\
        2  & L1AS6  & H-KYRRRR-OH     & 23 & L4AS6  & H-ASKRRR-OH      \\
        3  & L1AS7  & H-SKYRRRR-OH    & 24 & L4AS7  & H-RASKRRR-OH     \\
        4  & L1AS8  & H-ASKYRRRR-OH   & 25 & L4AS8  & H-SRASKRRR-OH    \\
        5  & L1AS9  & H-RASKYRRRR-OH  & 26 & L4AS9  & H-YSRASKRRR-OH   \\
        6  & L1AS10 & H-SRASKYRRRR-OH & 27 & L4AS10 & H-RYSRASKRRR-OH  \\
        \hline
        7  & L2AS4  & H-SRRR-OH       & 28 & L5AS4  & H-YRRR-OH        \\
        8  & L2AS5  & H-RSRRR-OH      & 29 & L5AS5  & H-AYRRR-OH       \\
        9  & L2AS6  & H-ARSRRR-OH     & 30 & L5AS6  & H-RAYRRR-OH      \\
        10 & L2AS7  & H-YARSRRR-OH    & 31 & L5AS7  & H-SRAYRRR-OH     \\
        11 & L2AS8  & H-RYARSRRR-OH   & 32 & L5AS8  & H-SSRAYRRR-OH    \\
        12 & L2AS9  & H-SRYARSRRR-OH  & 33 & L5AS9  & H-RSSRAYRRR-OH   \\
        13 & L2AS10 & H-KSRYARSRRR-OH & 34 & L5AS10 & H-KRSSRAYRRR-OH  \\
        \hline
        14 & L3AS4  & H-YRRR-OH       & 35 & L6AS4  & H-ARRR-OH        \\
        15 & L3AS5  & H-RYRRR-OH      & 36 & L6AS5  & H-RARRR-OH       \\
        16 & L3AS6  & H-SRYRRR-OH     & 37 & L6AS6  & H-SRARRR-OH      \\
        17 & L3AS7  & H-ASRYRRR-OH    & 38 & L6AS7  & H-YSRARRR-OH     \\
        18 & L3AS8  & H-RASRYRRR-OH   & 39 & L6AS8  & H-RYSRARRR-OH    \\
        19 & L3AS9  & H-SRASRYRRR-OH  & 40 & L6AS9  & H-KRYSRARRR-OH   \\
        20 & L3AS10 & H-KSRASRYRRR-OH & 41 & L6AS10 & H-SRKRYSRARRR-OH \\
        \hline
    \end{tabular}
\end{table}

\section{Catch22 Feature Descriptions}

\begin{table}[H]
    \caption{Description of the 11 catch22 features with the highest SHAP values.}
    \begin{tabular}{cc}
    Input & Description \\
    \hline
    \verb|acf_timescale| & The first intersection of the ACF with $1/e$.\\
    \verb|acf_first_min| & The first minimum of the ACF.\\
    \verb|ami2| & The mutual information of the autocorrelation with delay $\tau=2$.\\
    \verb|stretch_high| & The longest successive sequence greater than the mean.\\
    \verb|periodicity| & The first peak in the ACF.\\
    \verb|ami_timescale| & The minimum of the automutual information function. \\
    \verb|entropy_pairs| & The entropy of pairs after three-symbol-binning the signal.\\
    \verb|forecast_error| & The error of a simple 3-point rolling mean forecast.\\
    \verb|mean| & The mean of the signal.\\
    \verb|SD| & The standard deviation of the signal. \\
    \verb|log_length| & The (natural) logarithm of the dwell time. \\
    \end{tabular}
\end{table}

\subsection{Expression and Purification of Nanopores}
Biosynthesis of wt-pAeL was carried out in BL21(DE3)pLysS competent cells after transformation with the corresponding pET22b(+)::plb::pAeL-His6 construct and cultivation at \SI{37}{\celsius}. Gene expression was induced at an OD600 of 0.6 by adding \SI{500}{\micro\molar} IPTG and verified by SDS- PAGE analysis. 
Cells were further incubated at \SI{18}{\celsius} and harvested by centrifugation (\SI{1}{h}, \SI{4400}{rpm}, \SI{4}{\celsius}) when no further change in OD600 was observed. 
Cells were washed with cold PBS and pelleted by centrifugation (\SI{45}{min}, \SI{4400}{rpm}, \SI{4}{\celsius}). 
Cells were disrupted by ultrasonication, and debris was removed by ultracentrifugation (\num{200,000} xg, \SI{2}{h}, \SI{4}{\celsius}). 
Protein purification was carried out by Ni affinity chromatography (AFI), followed by size-exclusion chromatography (SEC) using an {\"A}KTA HPLC purifier system (GE Healthcare/Amersham Bioscience, Chalfont St Giles, Great Britain) equipped with a \SI{5}{mL} HisTrap HP or a Superdex 200 Increase 10/300 GL column. 
Proteins from the AFI elutions were pooled and concentrated to a final volume of \SI{250}{\micro\litre} prior to SEC using Amicon Ultra \SI{15}{\milli\litre} centrifuge filters with a molecular weight cutoff of \SI{10}{\kilo\dalton} (Merck KGaA, Darmstadt, Germany). 
The purity of the SEC elutions was verified by SDS-PAGE, and protein concentration was determined using a Bradford assay (Coomassie Brilliant Blue G-250, BioRad, Hercules, CA). 
Final protein samples were adjusted to \SI{1}{\milli\gram\per\milli\litre}, nitrogen shock-frozen, and stored at \SI{-80}{\celsius}. 
The activation of pAeL was carried out immediately before the experiment by applying trypsin (Promega, Madison, WI) at room temperature for 15 minutes.

\subsection{Electrophysiological Data Recording}
Nanopore recordings were executed using a modified Orbit16 platform (Nanion Technologies, Munich, Germany) in conjunction with MECA16 microelectrode cavity arrays (Ionera Technologies, Freiburg, Germany), following previously established protocols~\citep{ouldali2020, piguet21a, ensslen22a, martinez15a}. 
Specifically, one of the 16 coplanar gold lines connected to the MECA 16 cavities (\SI{50}{\micro\metre} diameter) was linked to the headstage of an Axopatch 200B patch clamp amplifier (Molecular Devices, Sunnyvale, CA) via a \SI{1}{cm} unshielded silver wire. 
The amplifier was configured in capacitive feedback mode with a \SI{50}{\milli\volt\per\pico\ampere} gain, and the internal low-pass filter was set to a \SI{100}{kHz} cutoff (\SI{-3}{\decibel}). The signal output was routed through an external low-pass Bessel filter (npi electronic, Tamm, Germany, 8- pole, custom LHBF-48X-8HL) with a \SI{50}{kHz} cutoff and digitized at a sampling rate of \SI{1}{MHz} using a PCI-6251 16-bit ADC interface (National Instruments, Austin, TX) managed by GePulse software (Michael Pusch, University of Genoa, Italy).
Electrodes were immersed in \SI{150}{\micro\litre} of \SI{4}{M} KCl, and lipid bilayers formed across the orifices of the MECA16 by utilizing an advanced bubble painting technique. 
Stable single AeL nanopores were obtained in most membranes within 10 minutes after the addition of activated pAeL. 
Peptides were synthesized via solid-phase custom synthesis (Intavis Peptide Services GmbH, T{\"u}bingen, Germany) and acquired as a lyophilized powder. 
The material was dissolved in \SI{5}{mM} HEPES, pH \num{7.5} and used at a final concentration of \SI{1}{\micro\molar}. 
Subsequently, the recorded data were exported in ABF format using GePulse for machine learning analysis.

\section{Labeling}

\begin{figure}[H]
    \centering
    \includegraphics{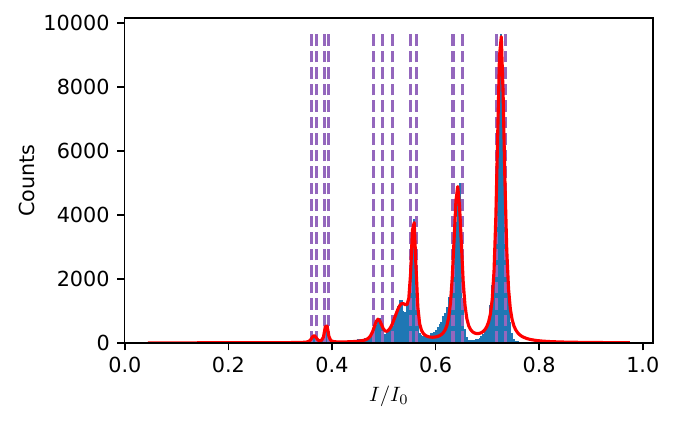}
    \caption{Voigt peaks fitted on the histograms of the normalized event means for ladder 1.}
\end{figure}

\begin{figure}[H]
    \centering
    \includegraphics{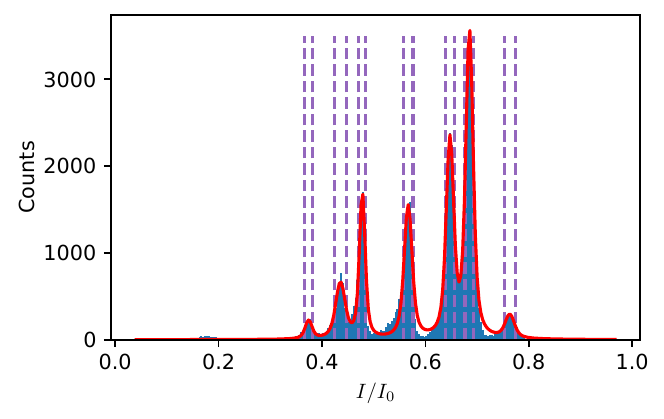}
    \caption{Voigt peaks fitted on the histograms of the normalized event means for ladder 2.}
\end{figure}

\begin{figure}[H]
    \centering
    \includegraphics{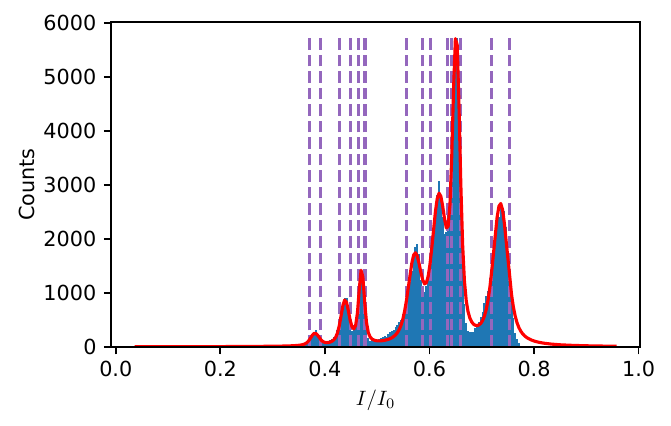}
    \caption{Voigt peaks fitted on the histograms of the normalized event means for ladder 3.}
\end{figure}

\begin{figure}[H]
    \centering
    \includegraphics{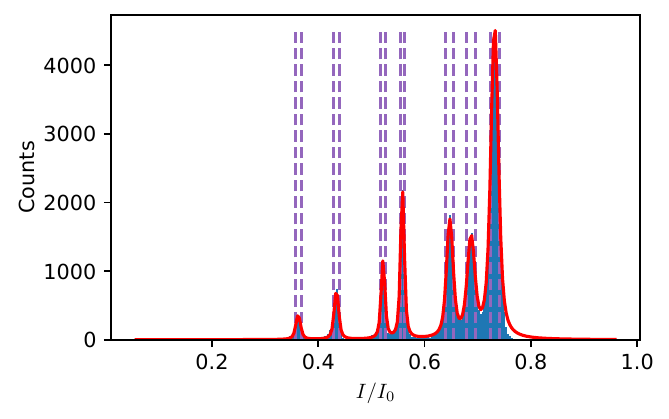}
    \caption{Voigt peaks fitted on the histograms of the normalized event means for ladder 4.}
\end{figure}

\begin{figure}[H]
    \centering
    \includegraphics{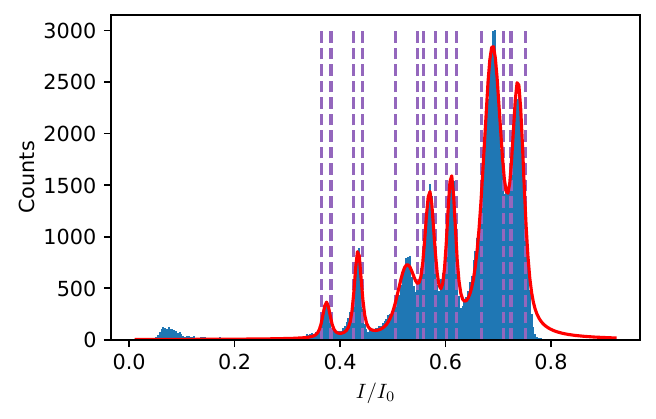}
    \caption{Voigt peaks fitted on the histograms of the normalized event means for ladder 5.}
\end{figure}

\begin{figure}[H]
    \centering
    \includegraphics{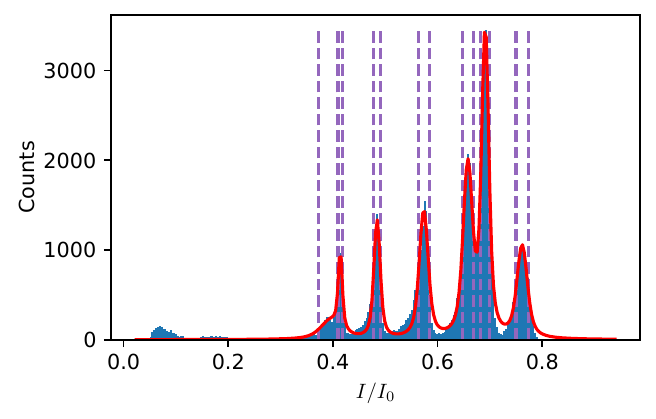}
    \caption{Voigt peaks fitted on the histograms of the normalized event means for ladder 6.}
\end{figure}

\bibliography{bibliography}

%% file: content/abstract.tex
Protein characterization using nanopore-based devices promises to be a breakthrough method in basic research, diagnostics, and analytics.
Current research includes the use of machine learning to achieve this task.
In this work, a comprehensive statistical analysis of nanopore current signals is performed and demonstrated to be sufficient for classifying up to 42 peptides with over 70 \% accuracy.
Two sets of features, the statistical moments and the \emph{catch22} set, are compared both in their representations and after training small classifier neural networks.
We demonstrate that complex features of the events, captured in both the catch22 set and the central moments, are key in classifying peptides with otherwise similar mean currents.
These results highlight the efficacy of purely statistical analysis of nanopore data and suggest a path forward for more sophisticated classification techniques.

%% file: content/introduction.tex
\section{Introduction}
\label{sec:introduction}

In the last decade, nanopore technology has emerged as a promising approach to third-generation DNA sequencing.
In nanopore sensing, an analyte is driven through a nanoscale pore while a current is measured between the entrance and the exit~\citep{wang21a}.
The presence of the analyte inside the pore leads to a drop in the current that, ideally, is characteristic of the analyte present, making it suitable for analyte characterization.
With its ability to produce long continuous reads in real-time and at low cost, the method has found a foothold in many areas of biomedicine, such as disease monitoring~\citep{hoenen2016, kafetzopoulou2019}, human genome research~\citep{jain2018} and tumour detection~\citep{vermeulen2023}.
Inspired by these successes, research has now begun to tackle the task of nanopore-based protein sensing in the rapidly growing field of proteomics, which would greatly benefit from new technology for efficient proteome exploration.

A peptide is defined as a chain of covalently bonded amino acids~\citep{hamley20a}, which, if large enough, is considered a protein.
During this work, we are interested in classifying peptides.
Classification, differs, however, to the sequencing often discussed in DNA.
DNA sequencing is made possible through the use of DNA polymerase, a substance that ratchets DNA through a pore one base pair at a time, thus allowing for a controlled readout of the sequence.
This approach is well established, and commercial solutions are available~\citep{wang21a}.
Machine learning has been applied extensively in improving the accuracy of these readouts~\citep{carral21, bao21, senanayake23, wang24, zhang24}.
However, such a mechanism is yet to be found in the case of peptides.
Instead, current research focuses on classifying the entire analyte as it moves into the pore.
Classification of peptides is of great interest to the medical community, in particular the identification of post-translational modifications, which have been shown to, among other things, indicate the presence of certain cancers~\citep{li21a, wang23a, srivastava23a, dutta23a}.

Several challenges underscore the complexity of nanopore-based peptide classification in comparison to DNA sequencing, among other things arising from a greater number of building blocks (amino acids) that constitute peptides.
This makes sequencing based on current signals more complex, albeit in principle possible for the discrimination on a single amino acid~\citep{ouldali2020}, or the detection of post-translational modifications~\citep{ensslen22a, cao24a}.
Recent studies demonstrated that aerolysin nanopores are also suitable for protein fingerprinting based on the different blocking currents of the peptide fragments of cleaved proteins~\citep{afsharbakshloo2022}.
As the peptide enters the pore, the current is reduced depending on the geometry of the polymer entering along with its interactions between the solvent, itself, and the pore walls.
As this blockage current depends on the geometry and interactions of the peptide, it can also be used to identify it.
While this process is highly non-trivial, it does offer a pathway for classifying these peptide fragments and, perhaps, eventually, entire proteins.
Similar reasoning has been used in work to predict properties of peptides based on the amino acid sequence and the structure of the peptide~\citep{badrinarayanan24a}.
In this work, however, we investigate the readouts of 42 distinct peptides driven through a wild-type aerolysin nanopore~\citep{ensslen24} to determine what methods can be used to classify them.

We explore using the statistical moments of the distributions formed by individual pore-entry events and more advanced statistical measures to build a representation of the events and use these representations to train simple neural network classifiers.
High-order moments of the distributions and time correlations in the signals are key in differentiating peptides with similar mean-blocking currents.
This work demonstrates that the classification of large numbers of peptides in a single sample is possible, paving the way for applications eventually in pathology or, in the near term, aiding scientists performing peptide-nanopore experiments.

%% file: content/results-and-discussion.tex
\section{Results and Discussion}
\label{sec:results-and-discussion}
\subsection{Feature Extraction}

\begin{figure}
    \centering\includegraphics{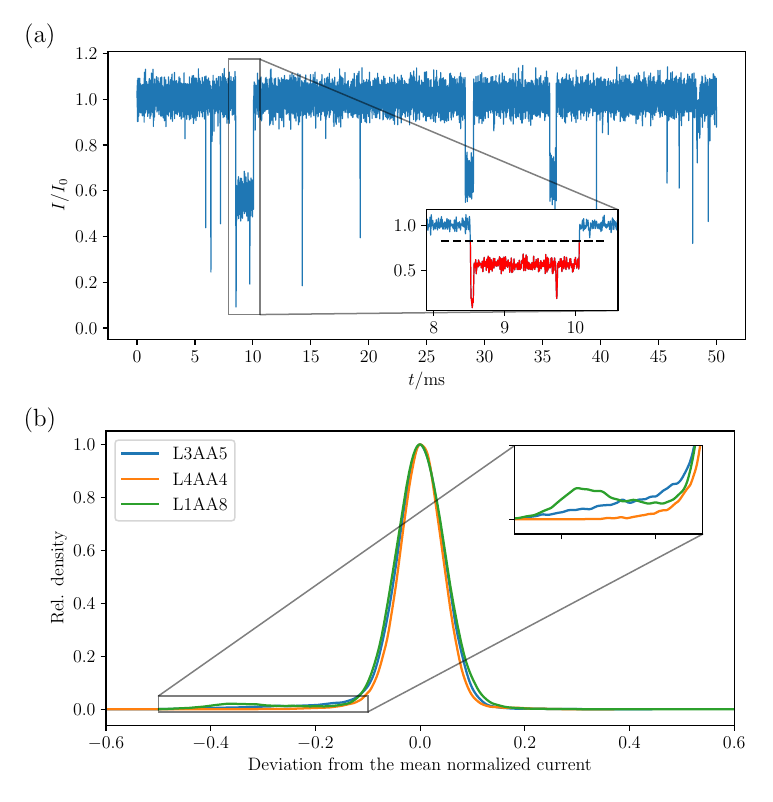}
    \caption{(a) Example of a normalized current readout with several events visible.
    The dashed line in the inset shows the cutoff, below which events are detected.
    The event current is shown in red.
    Details on the event detection and normalization are found in the methods section.
    (b) Histogram of the continuous signal of all events for the peptides L3AA5, L4AA4 and L1AA8. Each peak is shifted to be centered around 0 in order to visualize differences in the tails. 
}
\label{fig:feature-extraction}
\end{figure}

Starting from the current measurement, resulting in a time-series signal, as seen in \hyperref[fig:feature-extraction]{Figure 1(a)}, we perform an event detection, followed by a normalization with respect to the open pore current, as discussed in the methods section.
To illustrate that there are differences in the statistics of the resulting events of different peptides, we concatenate the normalized currents of all events of the peptides L3AA5, L4AA4 and L1AA8 into one continuous current signal and produce a histogram, respectively. Overlaying the resulting histograms by centering each peak around zero, one can identify differences in the width as well as other structural differences in the distribution, as illustrated in \hyperref[fig:feature-extraction]{Figure 1(b)}.
We aim to extract these differences in order to identify the individual peptides by their events by using two methods, which generate a fixed number of features of a time series to be used as feature vectors:
The set of the first 10 central moments, which will be called the moment decomposition set, and the catch22 feature set~\citep{lubba19a}.
Details on each feature set can be found in the methods section.

\begin{figure}
\centering
\includegraphics{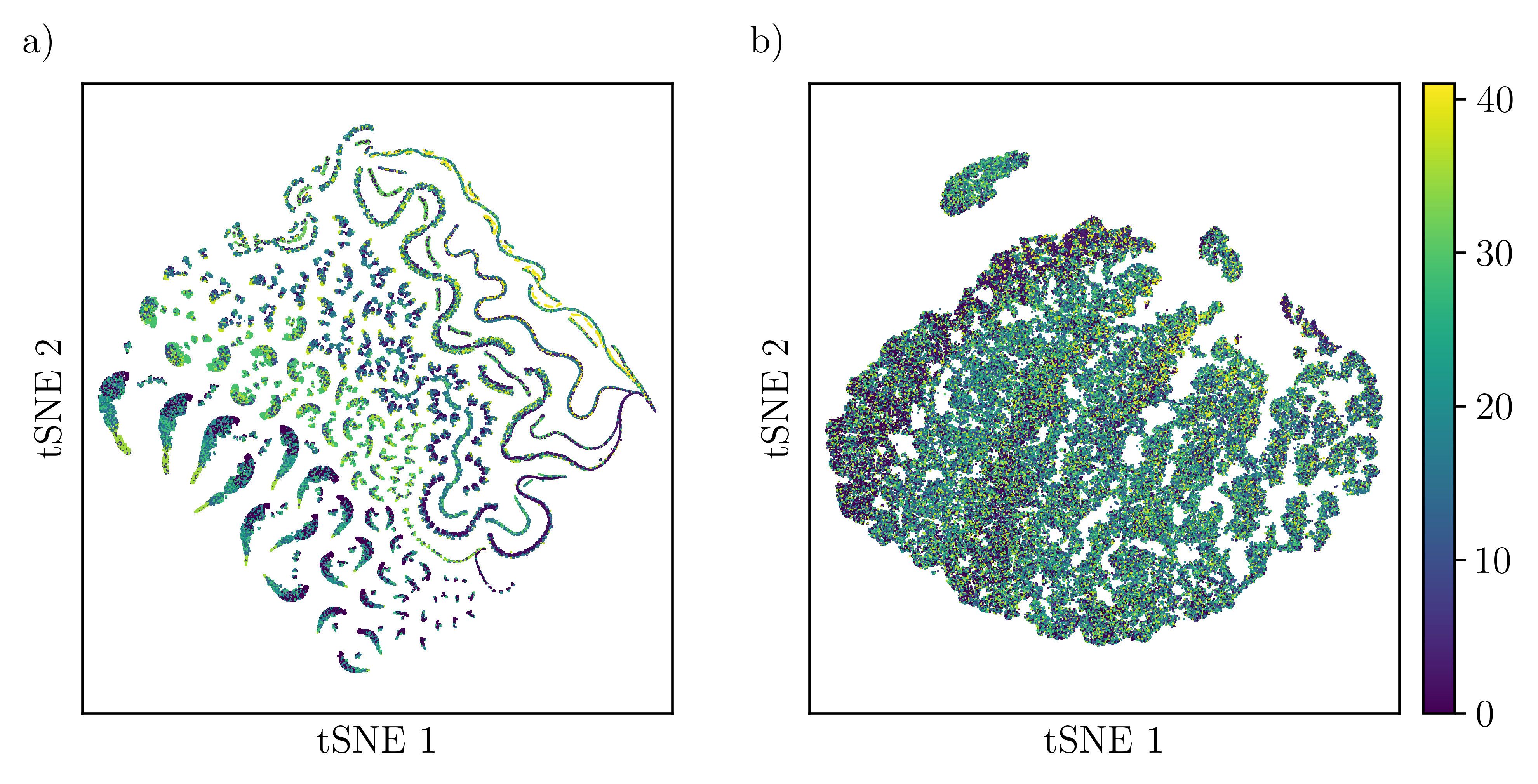}
\caption{t-SNE embedding of the two feature extraction methods studied here, moment decomposition (a) and catch22 (b).
Color represents the respective class associated with the coordinate. It is clear from the t-SNE plots that some structure is present in the vector representations. 
    However, it is not enough to make accurate statements regarding classification.}
    \label{fig:tsne_comparison}
\end{figure}

Before exploring the use of sophisticated classification algorithms for this, we perform a dimensionality reduction on the feature vectors to determine whether any structure can be found without additional processing.
\hyperref[fig:tsne_comparison]{Figures 2(a) and 2(b)} show a t-distributed Stochastic Neighbor Embedding (t-SNE)~\citep{vandermaaten08a} as implemented in the scikit-learn~\citep{pedregosa11a} (Additional t-SNE plots with varying parameters can be found in Figures 1 and 2 in the supplementary information.).
T-SNE embeds data from higher dimensions into lower dimensions while retaining their structure as much as possible.
This means that similar points in high-dimensional space will be mapped closer together in low-dimensional space.
It is a valuable tool for understanding structure in otherwise hard-to-visualize datasets.
In the ideal case, t-SNE embedding would represent the data from each featurization process into N distinct clusters for N classes, i.e. 42 in this work, each representing an individual peptide.
The datasets have a structure in both the moment decomposition and the catch22 feature extraction.
However, these clusters are not clear enough to identify 42 distinct classes.
Therefore, more advanced classification means must be used to distinguish events further.

\subsection{Model training}
\begin{figure}
    \centering\includegraphics{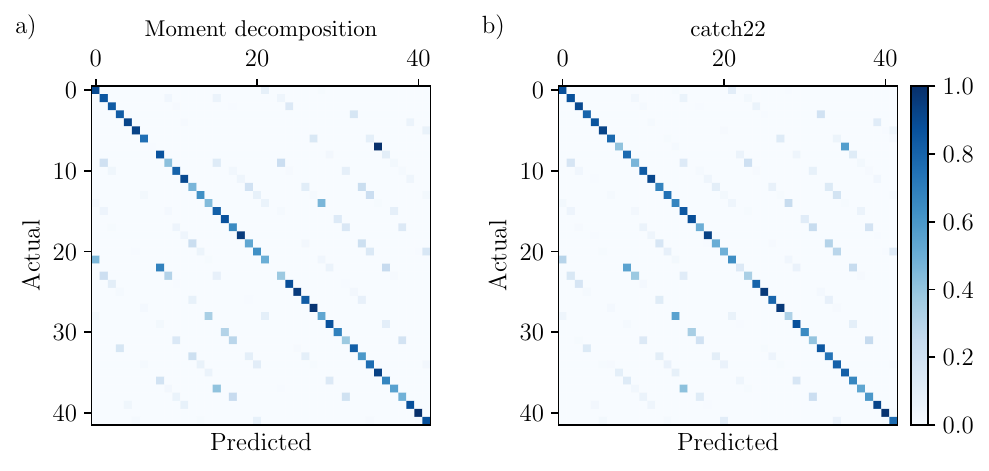}
    \caption{(a-b) Confusion matrices generated from the final models trained on the moment decomposition (a) and catch22 (b) datasets.}
    \label{fig:nn-results}
\end{figure}

The previous investigation established that the two statistical means of converting the pore-entry events into feature vectors yield a representation with intrinsic structure.
However, this structure alone is insufficient to separate classes; therefore, a more sophisticated means of classification is applied using the feature vectors as inputs, namely, feed-forward neural networks.
Further information regarding the network training can be found in the methods section.
Figure~\ref{fig:nn-results} shows the results of training neural networks on the moment decomposition and catch22 data sets.
In Figure~\ref{fig:nn-results}, the confusion matrices from the moment decomposition and catch22 trained models, respectively, are shown.
The confusion matrices provide an overview of the classification ability of the tested models.
The rows correspond to the peptide predicted by the neural network, whereas the columns correspond to the actual peptide label assigned to the data.
Rows in the matrix are normalized by the true labels as to avoid potential bias due to data imbalance.
In the ideal case, these matrices have a solid diagonal color and no off-diagonal predictions.
In both cases, this holds only for a limited number of peptides, but overall, both methods show strong predictive performance.
Table~\ref{tab:metrics} outlines additional metrics for understanding these results further.

\begin{table}
\caption{
Classification results of the dense neural networks after training on each feature set.
The precision, recall and $F_1$-score computed by weighting each class metric by the number of available labels of that class.
}
\begin{tabular}{c|cccc}
Feature set & Balanced accuracy & Precision & Recall & $F_1$-score\\
\hline
Moment decomposition & \num{0.728} & \num{0.731} & \num{0.730} & \num{0.726} \\
Catch22 & \num{0.736} & \num{0.737} & \num{0.736} & \num{0.733}
\label{tab:metrics}
\end{tabular}
\end{table}

\begin{figure}
    \centering
    \includegraphics[width=1\linewidth]{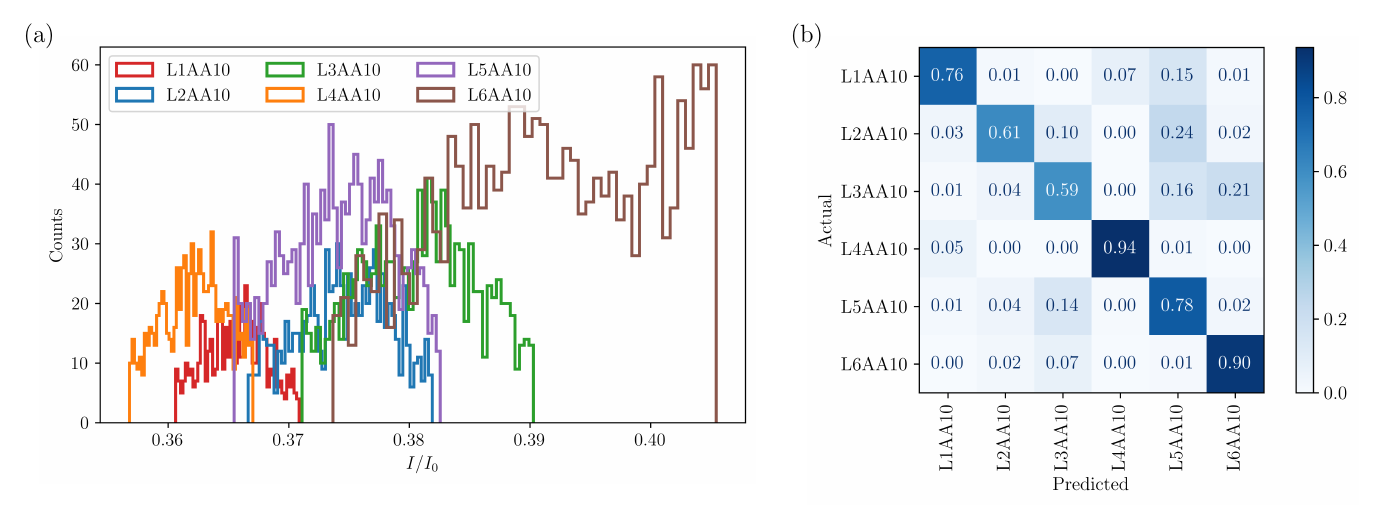}
    \caption{(a) Histogram of the mean blocking currents of the decapeptides of each peptide ladder.
    Labelling was performed based on an interval only around each peak, which explains the cut sharp edges of all histograms.
    (b) Model results of the moment decomposition model, tested on the decapeptide samples in the test dataset.}
    \label{fig:decapeptides}
\end{figure}

Table~\ref{tab:metrics} show that the catch22 trained model achieved marginally better classification performance than the moment decomposition, although it still failed to discern all peptide ladder classes fully.
The precision score, which indicates how often a peptide ladder was correctly predicted compared to the total number of times it was predicted, is slightly improved in the catch22 model, suggesting a smaller number of false predictions on average.
The recall scores show a similar trend, with the catch22 trained models less likely to miss a true classification.
Therefore, it is unsurprising that the $F_{1}$-score, comprised of both precision and recall, also follows this trend.
This relatively simple analysis method can classify the 42 peptide classes surprisingly well.
However, in many of these cases, mean current alone can be used, as outlined in~\citet{behrends2022}.
Therefore, to highlight the specific role of the higher-order moments, the model capabilities of distinguishing between the six decapeptides is shown in Figure~\ref{fig:decapeptides}.
These decapeptides are chosen as they have identical mass, resulting in a high degree of overlap in the event mean histogram, making prediction based on the mean alone infeasible.
Specifically, the confusion matrix in Figure~\ref{fig:decapeptides}b shows that identification of the decapeptides can be achieved beyond the straightforward comparison of the mean current of the normalized events, highlighting that additional information lies in higher-order moments of the signal, and supporting the use of more sophisticated classification methods.
Note that the histograms in Figure~\ref{fig:decapeptides}a do not represent the optimally achievable identification of the mean current.
Rather, the event detection explicitly includes parts of the rising and falling phase in each event, which broadens and possibly skews all peaks. 
This can be accounted for by employing additional steps in the event preprocessing~\citep{ensslen24}.
For this task however, we explicitly include the rise and fall of each event based on the assumption that they might also contribute to the features in both feature sets.

By demonstrating that we can classify peptides with highly overlapping current signals, we show that in the future, such a method could be used to analyze large samples of peptides without extensive preprocessing, and that perhaps separation of these peptides in a lab setting can also be avoided.
This opens not only open new possibilities for scientists performing nanopore experiments, but also provides a means for future investigation into clinical peptide classification applications, where extensive distillation of peptides before classification might be expensive or impractical.

\subsection{SHAP Model Explanation}
\begin{figure}
    \centering
    \includegraphics[width=\linewidth]{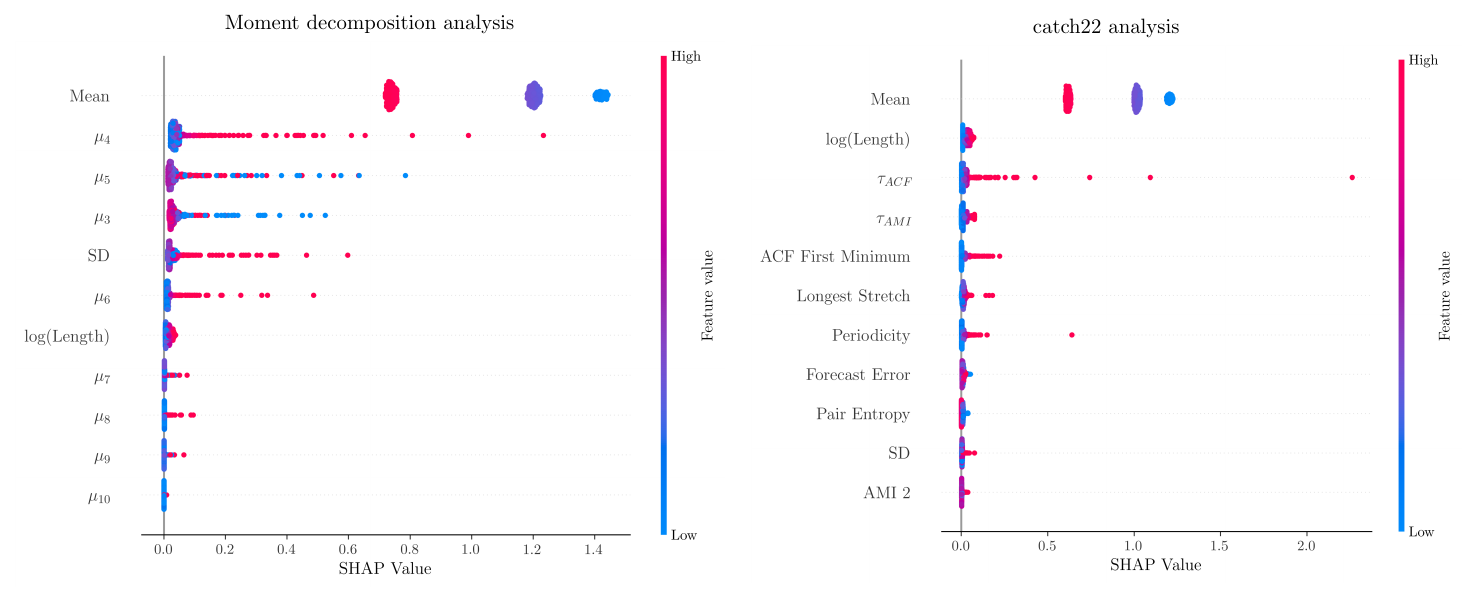}
    \caption{SHAP analysis for both the statistical moments (left) and catch22 (right) computed over all 42 peptides. The top 11 features were taken from both, and the rest fell below statistical significance. A detailed overview of the catch22 features can be found in Table 1 of the supplementary information.}
    \label{fig:shap}
\end{figure}

Aside from simplicity, an additional benefit of using the model inputs chosen here is their interpretability.
Specifically, the variables being passed to the network are well-defined measurements.
By performing SHAP analysis on the trained models, we can thus evaluate the importance of all features in each feature set.
The results of this analysis are shown in Figure~\ref{fig:shap}.
In a standard SHAP figure, the SHAP value indicates whether the variable increased (positive) or decreased (negative) the model output.
The color of each point indicates whether a large (red) or small (blue) scalar value of that feature, the feature value, led to the change.
For example, a small feature value with a large SHAP score would indicate that small scalar values for that feature were correlated with larger model outputs.
To study the impact of the features with respect to their scalar values, the absolute SHAP values are shown.
Examining Figure~\ref{fig:shap}, several interesting outcomes are evident.
First, input features present in both datasets, such as the mean blocking current and the logarithmic dwell time, show similar importance characteristics.
The catch22-trained models also seemed to depend greatly on the autocorrelation time as estimated by computing the autocorrelation function of an event, indicating that dynamic properties also have an impact on the model prediction. 
This was completely missed in the statistical moment analysis but might be key to further improvements in predictive performance.

Not taken into account are correlations between features, which was not captured by the SHAP analysis.
However, the similarity in importance scores of the inputs present in both datasets suggests that the correlations of inputs are negligible.

%% file: content/conclusion.tex
\section{Conclusions}
\label{sec:conclusion}
We have demonstrated that the moment decomposition of the time series data generated in nanopore experiments, either through the moments of the distributions formed by the events or more sophisticated analysis with the catch22 feature set, can be used to construct feature vectors for input into classification models.
Based on the non-Gaussian form of the histograms, these features capture information present in the higher order moments of the current signal distributions.
We showed that using small neural networks, classification accuracies of 72.8 \% and 73.6 \% can be achieved for the moments and catch22, respectively, with up to 42 peptide classes.
These results were emphasized by showing the performance on the decapeptides where single event distinction is hindered by overlaps in the relative current mean histogram.
We further found through SHAP analysis that the signal's dynamic properties yielded additional classification performance in the catch22 dataset and should be considered in more complex feature decomposition approaches.
This simple analysis is computationally efficient and capable of previously unseen classification performance on many classes.
Our results highlight the role machine-learning classification can play in classification of large numbers of peptides that may one day be used in a clinical setting.
Further, we demonstrate that preparative separation of peptides in experimental settings may not be necessary, as our models can differentiate those with highly overlapping mean blockade currents.

%% file: content/methods.tex
\section{Methods}
\label{sec:methods}
\subsection{Peptides}\label{subsec:peptide-ladders}
\begin{figure}
\input{content/peptide-ladder}
\caption{Description of peptide ladders used in the investigation.}
\label{fig:peptide-ladders}
\end{figure}
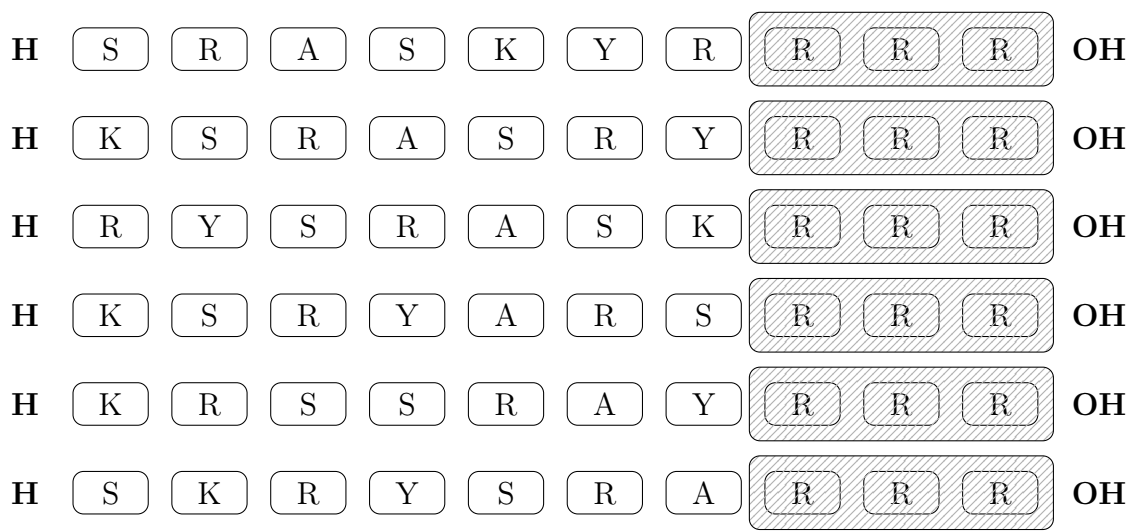

In this investigation, we perform classification on 42 peptides, successively constructed by one to six N-terminal amino acids.
Figure~\ref{fig:peptide-ladders} demonstrates the construction of the 6 individual ladders taken from~\citet{behrends2022}.
The far left side shows the N-terminus with the hydrogen group, and the far right shows the C terminus with the hydroxide group. Letters S, R, A, K, and Y correspond to the amino acids Serine, Arganine, Alanine, Lysine, and Tyrosine respectively.

\subsection{Event Detection and Labeling}
\label{subsec:event-detection-and-labeling}
During nanopore experiments, a time series of current measurements is captured, within which many events will occur.
Before these events can be classified, they must first be isolated from this large sequence.
In this work, a statistical threshold detection approach\citep{piguet21a, cao24a} was used, wherein deviations from the Gaussian-like distribution of the open pore current are used to detect outliers.
If a current was detected below three standard deviations, $\sigma$, from the expected open-pore mean current, it was considered the beginning of an event.
To avoid unwanted events in the dataset, a wavelet threshold filter using the \verb|bior1.5| with a hard threshold of $0.5$ was applied before the detection algorithm, and events whose mean current was above a second threshold of $4\sigma$ were ignored in line with~\citet{piguet21a}. 
Further reduction of spurious events was achieved by only allowing events with a dwell time greater than \SI{80}{\micro\second}.
Finally, all detected events were normalized by the mean of the surrounding open pore current.
To account for drift in open pore current over time, a consequence of the experimental configuration being used, the mean open pore current was calculated by taking a window around each detected event and ignoring other events that might fall within this window.
For labeling, the peaks in the histogram were fitted with Voigt functions.
Peptide-specific events were then identified if they fell within the full-width-half-maximum (FWHM) of each peak.
Strictly speaking, this discrimination of events for labeling poses a potential source for errors, as neighboring histograms overlap with  each other and may, therefore, fall within the FWHM of their neighbor, thus ending up falsely labelled.
Since labeling is done using the normalized event mean histogram for all measurements of one specific ladder, there are no labeling errors between peptides with the same amount of amino acids.
We therefore expect the amount of data leakage in our approach to be small.
The plots used for labelling can be found in the supporting information.

\begin{figure}
    \centering
    \includegraphics[width=\linewidth]{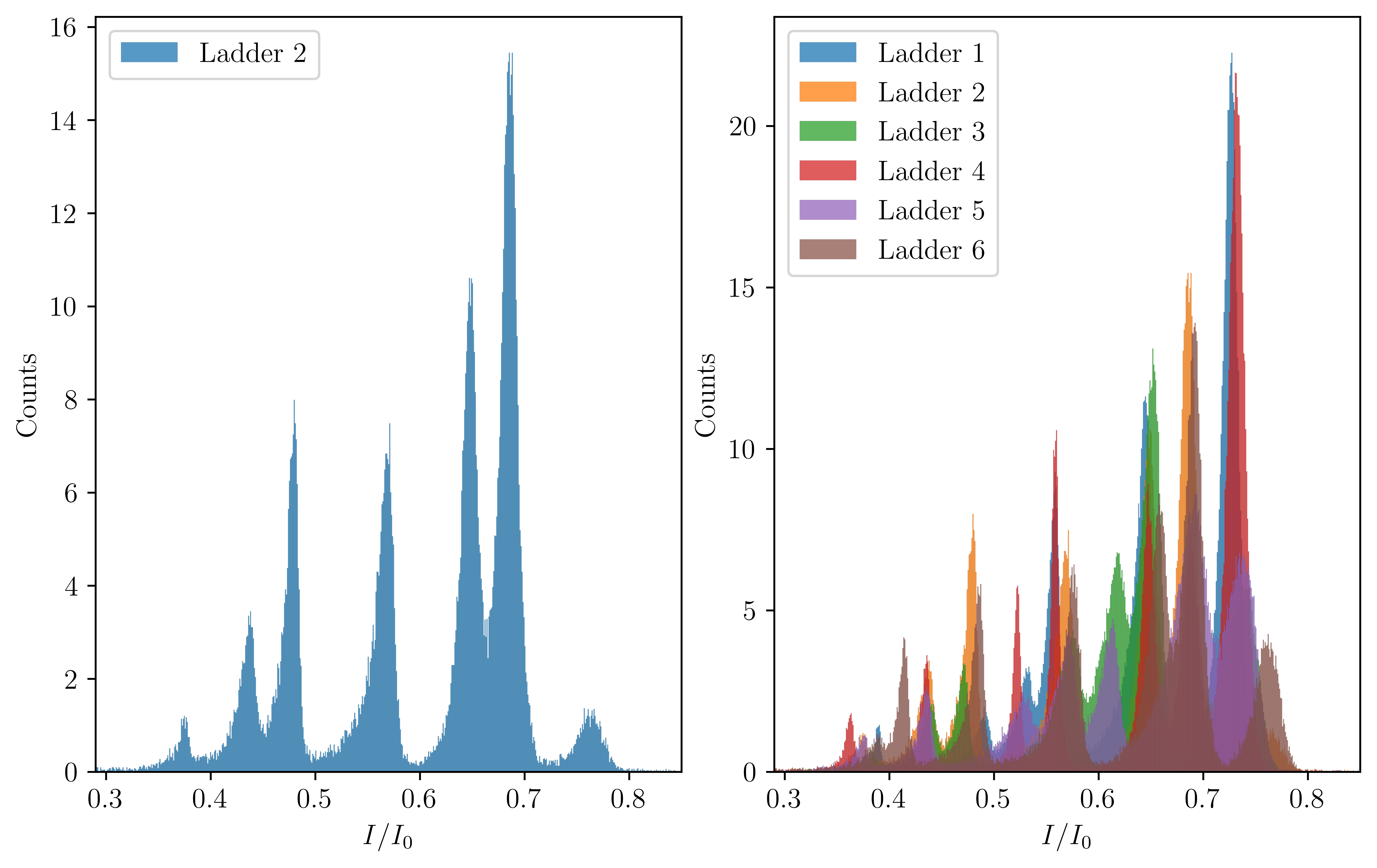}
    \caption{Normalized mean event histograms for all 42 peptides considered in a single experiment. (left) Histogram formed from the normalized event mean of the L2 ladder experiment. (right)Histogram formed from the combined measurements of all 6 ladders.}
    \label{fig:all-peptides}
\end{figure}

During data collection, each peptide, along with all of its subchains, are studied in a single experiment.
This separation of chains allows for easier labelling as there is less overlap in the current amplitude histograms, an example of which is shown in Figure~\ref{fig:all-peptides}.
However, in the classification task, we treat the data as though all 42 peptide chains are being measured together in a single sample, resulting in the amplitude histograms shown in Figure~\ref{fig:all-peptides} and making the use of more sophisticated machine-learning and statistical feature extraction necessary.

\subsection{Moments Analysis}
\label{subsec:moment-analysis}
The $n^{th}$ central moment of a random variable, $X$, is defined as
\begin{equation}
    \mu^{n} = \langle (X - \mu)^{n} \rangle = \sum\limits_{i} (x_{i} - \mu)^{n} f(x_{i}),
\end{equation}
for probability density $f$ and mean $\mu$~\cite{papoulis02a}.
The set of all moments for a distribution can completely define it and comprise a complete or overcomplete representation, making them useful classification problems.
In this work, the distribution formed by computing the kernel density estimate~\citep{rosenblatt56a, parzen62a} is used to extract the first ten moments of the distributions of magnitudes of current signals, which form a feature vector for each pore-entry event.
The moments themselves were computed using Scipy~\citep{virtanen20a}.
In addition to these moments, we also included the natural logarithm of the dwell time of the events in the feature vector, resulting in 11 final features for each event.

\subsection{Catch22 Feature Extraction}
\label{subsec:catch22-feature-extraction}
Another approach to feature extraction performed in this study is using the \textit{catch22} framework from~\citet{lubba19a} to generate feature vectors.
catch22 is constructed from the larger HCTSA feature extraction toolkit~\citep{fulcher13a, fulcher17a}, which produces more than seven thousand features of a time series.
In their 2019 work,~\citet{lubba19a} demonstrated that these seven thousand features can be reduced to twenty-two with minimal performance loss when used for classification tasks.
The feature extraction is performed in this work using the Python package provided by the initial research group.
The extracted features, including the mean current, the standard deviation and the natural logarithm of the dwell time, are used in further analysis.

\subsection{SHAP Analysis}
\label{subsec:shap-analysis}
SHapley Additive exPlanation (SHAP) analysis~\citep{lundberg17a} is an approach built on game theory to explain how the outputs of a machine learning model are impacted by the inputs, in many ways, explaining what factors in the input data have led to the result.
It does so by using the SHAP value, a number that indicates how a certain input value (high or low) impacted the network's output by either increasing or decreasing it.
A large positive SHAP score associated with a large input number indicates that large values of this input increase the network's output.
The same logic carries for negative SHAP values.
SHAP analysis can be used to better understand the effect of the neural network inputs on predictive performance. 
This, in turn, simplifies problems by exposing redundant inputs and illuminates how certain conclusions are reached using the model.
This work utilizes the \texttt{shap} Python package to analyze the predictions generated by the neural networks.

\subsection{Neural Networks}
\label{subsec:neural-networks}
All neural networks were implemented using the PyTorch library~\citep{paszke19a}.
The architectures used in the final model fits are outlined in Table~\ref{tab:architecture} along with their relevant hyperparameters used for the training.
These architectures were chosen based on a parameter search over widths from 32 to 512 and depths from 1 to 5.
\begin{table}[]
\centering
\begin{tabular}{@{}l|l@{}}
\toprule
Architecture & $\text{ReLU}\left(\mathcal{D}^{64}\right)\text{ReLU}\left(\mathcal{D}^{128}\right)\text{ReLU}\left(\mathcal{D}^{42}\right)$ \\
Batch Size   & 1000                                                                                                                        \\
Optimizer    & AMSGrad(0.001)                                                                                                              \\
Train Data   & 242601                                                                                                                      \\ \bottomrule
\end{tabular}
\caption{Neural network parameters used in the statistical moments and catch22 dataset training.
ReLU activation functions are used between all dense layers, the dimension of which is specified by the superscript. 
The parameters are trained using the AMSGrad optimizer~\citep{reddi19a} with a learning rate 0.001.}
\label{tab:architecture}
\end{table}

\subsection{Confusion Matrices}
\label{subsec:confusion-matrices}
A confusion matrix for multiclass classification is an essential tool for evaluating the performance of a classification model. 
It visually represents the classifier’s predictions compared to the actual classes, with each row corresponding to actual peptide ladder classes and each column to predicted classes. 
Key metrics such as precision, recall, and F1 score can be derived from the confusion matrix. 
Precision measures the proportion of true positive predictions among all positive predictions, indicating the model's accuracy in identifying a specific peptide ladder class. 
Recall, or sensitivity, measures the proportion of true positives among actual positives, reflecting the model’s ability to capture all instances of a specific class. 
The F1 score, the harmonic mean of precision and recall, provides a balanced metric that considers false positives and negatives, making it particularly useful when dealing with imbalanced classes.

%% file: content/peptide-ladder.tex
\begin{tikzpicture}[
    ladder/.style={rectangle, draw, minimum size=1cm, font=\small},
    acid/.style={draw, rounded corners, minimum width=1cm, align=center},
    terminus/.style={font=\bfseries, anchor=west},
    faint/.style={draw, rounded corners, minimum width=1cm, align=center, fill=gray!20},
    checkered/.style={draw, rounded corners, minimum width=1cm, align=center, pattern=north east lines, pattern color=gray!60}
]

\node[terminus] (L1H) at (0,0) {H};
\node[acid, right=0.3cm of L1H] (L1S1) {S};
\node[acid, right=0.3cm of L1S1] (L1R) {R};
\node[acid, right=0.3cm of L1R] (L1A) {A};
\node[acid, right=0.3cm of L1A] (L1S2) {S};
\node[acid, right=0.3cm of L1S2] (L1K) {K};
\node[acid, right=0.3cm of L1K] (L1Y) {Y};
\node[acid, right=0.3cm of L1Y] (L1R2) {R};
\node[acid, right=0.3cm of L1R2] (L1R3) {R};
\node[acid, right=0.3cm of L1R3] (L1R4) {R};
\node[acid, right=0.3cm of L1R4] (L1R5) {R};
\node[checkered, fit={(L1R3) (L1R4) (L1R5)}, inner sep=0.2cm] (box) {};
\node[terminus, right=0.3cm of L1R5] (L1OH) {OH};

\node[terminus, below=0.6cm of L1H] (L2H) {H};
\node[acid, right=0.3cm of L2H] (L2K1) {K};
\node[acid, right=0.3cm of L2K1] (L2S1) {S};
\node[acid, right=0.3cm of L2S1] (L2R1) {R};
\node[acid, right=0.3cm of L2R1] (L2A1) {A};
\node[acid, right=0.3cm of L2A1] (L2S2) {S};
\node[acid, right=0.3cm of L2S2] (L2R2) {R};
\node[acid, right=0.3cm of L2R2] (L2Y1) {Y};
\node[acid, right=0.3cm of L2Y1] (L2R3) {R};
\node[acid, right=0.3cm of L2R3] (L2R4) {R};
\node[acid, right=0.3cm of L2R4] (L2R5) {R};
\node[checkered, fit={(L2R3) (L2R4) (L2R5)}, inner sep=0.2cm] (box) {};
\node[terminus, right=0.3cm of L2R5] (L2OH) {OH};

\node[terminus, below=0.6cm of L2H] (L3H) {H};
\node[acid, right=0.3cm of L3H] (L3S1) {R};
\node[acid, right=0.3cm of L3S1] (L3R) {Y};
\node[acid, right=0.3cm of L3R] (L3A) {S};
\node[acid, right=0.3cm of L3A] (L3S2) {R};
\node[acid, right=0.3cm of L3S2] (L3K) {A};
\node[acid, right=0.3cm of L3K] (L3Y) {S};
\node[acid, right=0.3cm of L3Y] (L3R2) {K};
\node[acid, right=0.3cm of L3R2] (L3R3) {R};
\node[acid, right=0.3cm of L3R3] (L3R4) {R};
\node[acid, right=0.3cm of L3R4] (L3R5) {R};
\node[checkered, fit={(L3R3) (L3R4) (L3R5)}, inner sep=0.2cm] (box) {};
\node[terminus, right=0.3cm of L3R5] (L3OH) {OH};

\node[terminus, below=0.6cm of L3H] (L4H) {H};
\node[acid, right=0.3cm of L4H] (L4S1) {K};
\node[acid, right=0.3cm of L4S1] (L4R) {S};
\node[acid, right=0.3cm of L4R] (L4A) {R};
\node[acid, right=0.3cm of L4A] (L4S2) {Y};
\node[acid, right=0.3cm of L4S2] (L4K) {A};
\node[acid, right=0.3cm of L4K] (L4Y) {R};
\node[acid, right=0.3cm of L4Y] (L4R2) {S};
\node[acid, right=0.3cm of L4R2] (L4R3) {R};
\node[acid, right=0.3cm of L4R3] (L4R4) {R};
\node[acid, right=0.3cm of L4R4] (L4R5) {R};
\node[checkered, fit={(L4R3) (L4R4) (L4R5)}, inner sep=0.2cm] (box) {};
\node[terminus, right=0.3cm of L4R5] (L4OH) {OH};

\node[terminus, below=0.6cm of L4H] (L5H) {H};
\node[acid, right=0.3cm of L5H] (L5S1) {K};
\node[acid, right=0.3cm of L5S1] (L5R) {R};
\node[acid, right=0.3cm of L5R] (L5A) {S};
\node[acid, right=0.3cm of L5A] (L5S2) {S};
\node[acid, right=0.3cm of L5S2] (L5K) {R};
\node[acid, right=0.3cm of L5K] (L5Y) {A};
\node[acid, right=0.3cm of L5Y] (L5R2) {Y};
\node[acid, right=0.3cm of L5R2] (L5R3) {R};
\node[acid, right=0.3cm of L5R3] (L5R4) {R};
\node[acid, right=0.3cm of L5R4] (L5R5) {R};
\node[checkered, fit={(L5R3) (L5R4) (L5R5)}, inner sep=0.2cm] (box) {};
\node[terminus, right=0.3cm of L5R5] (L5OH) {OH};

\node[terminus, below=0.6cm of L5H] (L6H) {H};
\node[acid, right=0.3cm of L6H] (L6S1) {S};
\node[acid, right=0.3cm of L6S1] (L6R) {K};
\node[acid, right=0.3cm of L6R] (L6A) {R};
\node[acid, right=0.3cm of L6A] (L6S2) {Y};
\node[acid, right=0.3cm of L6S2] (L6K) {S};
\node[acid, right=0.3cm of L6K] (L6Y) {R};
\node[acid, right=0.3cm of L6Y] (L6R2) {A};
\node[acid, right=0.3cm of L6R2] (L6R3) {R};
\node[acid, right=0.3cm of L6R3] (L6R4) {R};
\node[acid, right=0.3cm of L6R4] (L6R5) {R};
\node[checkered, fit={(L6R3) (L6R4) (L6R5)}, inner sep=0.2cm] (box) {};
\node[terminus, right=0.3cm of L6R5] (L6OH) {OH};

\end{tikzpicture}